\documentclass[twocolumn,showpacs,preprintnumbers,amsmath,amssymb,a4paper,superscriptaddress, nofootinbib,aps,pra,10pt]{revtex4-2}

\usepackage[utf8]{inputenc}  
\usepackage[main=english]{babel}  
\usepackage[T2A]{fontenc}

\usepackage{xcolor} 

\usepackage{float}
\usepackage{amsmath}
\usepackage{graphicx}
\usepackage{dcolumn}
\usepackage{bm}

\usepackage{latexsym}
\usepackage{amssymb}
\usepackage{eqnarray} 

\usepackage{mathrsfs}
\usepackage[percent]{overpic}

\begin{document}

\selectlanguage{english} 

\title{
Field-free alignment and orientation of linear molecules by two-color trapezoidal laser pulses
}
\author{Eugene A. Koval}
\email[]{e-cov@yandex.ru}
\affiliation{Bogoliubov Laboratory of Theoretical Physics, Joint Institute for Nuclear Research, Dubna, Moscow Region 141980, Russian Federation}

\date{\today}

\begin{abstract}\label{txt:abstract}
The field-free alignment and orientation of the linear molecule by the two-color trapezoidal laser pulses were  theoretically investigated. The trapezoidal shape of a laser pulse allows to enhance the maximum alignment degree for the same intensity and duration comparing to the conventional Gaussian laser pulse. The alignment and orientation persist after the pulse for both non-adiabatic and adiabatic regimes. While the maximum (during the pulse) alignment degree quickly saturates and remains almost constant with the pulse duration increase, the dependencies of the maximum (outside the laser pulse) alignment and orientation degrees on the pulse duration show the clear periodic structures in the adiabatic regime. The effect of the non-zero temperature is also shown. 
Applying additional the monochromatic or two-color prepulse increases the maximum orientation degree, but the application of the two-color prepulse leads to a higher maximum orientation degree than the monochromatic prepulse.
The effect of the relative phase variation on the molecular orientation in case of one and two pulses was also discussed.
\end{abstract}

\maketitle

\section{Introduction}
\label{sec:Intro}

Recently, a rapid growth of theoretical \cite{wang2020optimal, mun2019orientation, kanai2001numerical} and experimental \cite{mun2019orientation, zhang2011field, ohmura2004quantum, oda2010all} studies of the alignment and orientation in laser and combined fields have been seen~\cite{koch2019quantum, nautiyal2021orientation,hong2023quantumREVIEW}. The methods of the molecular alignment and orientation control are practically used in applications such as multi-photon ionization\cite{suzuki2004optimal}, angular distribution of photo-electrons~\cite{holmegaard2010photoelectron}, high harmonics generation~\cite{itatani2004tomographic,kraus2012high}, the operations on a qubit for quantum computer, based on the polar molecules in the optical lattices~\cite{mun2022all}.

The alignment and the orientation of neutral molecules under free of external static electric field  conditions depends on the many laser field parameters such as the intensities, the frequencies, the shapes and the duration times of laser pulses, time of delay between the pulses, etc.~\cite{koch2019quantum, nautiyal2021orientation,hong2023quantumREVIEW}.

Compared with molecular alignment, achieving molecular orientation is more challenging, since the degree of the orientation usually dramatically decreases with the temperature increase. There are many approaches proposed in order to enhance the orientation degree. The orientation enhancement by the presence of the static electric field~\cite{friedrich1999enhanced, sakai2003controlling} is applicable to the polar molecules, but it may perturb the quantum states of the molecule. 

Another used in a bunch of papers all-optical method is the creation of the asymmetric interaction potential by a two-color laser field, initially proposed by Kanai~et.al.~\cite{kanai2001numerical} and experimentally proven~\cite{de2009field, oda2010all, kraus2014two}. Higher values of the laser field intensity for this method, usually, lead to the higher alignment and orientation degree, but there always is an ionization limit, that restricts such a direct way of achieving better alignment and orientation and stimulates the optimization of the laser pulses' parameters~\cite{koch2019quantum}. The trains of one-color and two-color pulses with different shapes~\cite{leibscher2003molecular, bisgaard2004observation, lee2004phase, wu2010field, zhang2011controlling, ren2014alignment, tyagi2017effect} were investigated in order to enhance the quantum control of the molecular rotation, but avoid the ionization limit. The saturation of the alignment~\cite{leibscher2003molecular} was experimentally proven for the quantum-state-selected NO molecules by short pulse in Ref.\cite{ghafur2009impulsive}. The effects of the perpendicularly crossed polarizations and elliptical polarization of two-color pulses on the field-free orientation were studied in Ref.~\cite{mun2020field} and in Ref.~\cite{hossain2020all} respectively. The laser field with a slow turn on and rapid turn off schemes were shown to adiabatically induce the substantial degree of orientation in controllable manner~\cite{seideman2001dynamics,muramatsu2009field}. The complex optimal control schemes were also proposed to enhance~\cite{haj2002numerical, dion2005optimally} and to shape the time evolution~\cite{tehini2019shaping} of the degree of orientation. The recent experimental high-resolution imaging of the time-dependent angular probability density distribution of a rotational wavepacket~\cite{karamatskos2019molecular} showed, that a complete ``molecular movie'' of quantum rotation is now possible to observe directly.

The recent theoretical studies~\cite{kitano2011high,lapert2012field,hong2023quantumREVIEW}(and references therein) showed that a combined approach of the two-color non-resonant prepulse with the THz pulse is expected to highly increase the resulting orientation, while the fine adjustment of the pulses' parameters is still crucial. 
Although the THz pulses are theoretically efficient in molecules orienting, but there are still the problems with the available sources of THz radiation~\cite{fedorov2020powerful}. Experiments on orienting the room-temperature OCS molecules~\cite{fleischer2011molecular} (200 K) and the cooled HBr molecules (8 K)~\cite{kitano2013orientation} by the intense single-cycle THz pulse showed rather weak degree of orientation. 
Since the advantages of the THz pulses are seen mainly for to the resonant rotational excitation~\cite{hong2023quantumREVIEW, fleischer2011molecular}, there is a drawback, that the THz laser has to have a finely tuned frequency specific to the every molecule. Otherwise, as shown in recent theoretical studies (see Fig.2 of Ref.~\cite{ni2020study}) the maximum degree of orientation could have the quadrupled decrease.

The envelope shapes used in the papers mentioned above are mainly the Gaussian, $\sin^2$~\cite{nautiyal2021orientation}, square~\cite{zhang2011field} shapes. 
In our article,we investigate field-free molecular alignment by a two-color laser field with the trapezoidal shape of the pulse. Urvashi et al.~\cite{arya2013field} considered ``ramped'' (with $\sin^2$-rise, plateau and $\sin^2$-fall) shape of pulses, but only the two \textit{one-color THz} pulses were studied to orient the molecule by the interaction with the dipole moment and neglecting the polarizability and hyperpolarizability components due to the low intensities of the THz field. To the best of our knowledge, the application of the trapezoidal shape of the nonresonant two-color $(\omega,2\omega)$ laser pulse to the alignment and orientation enhancement was not studied before, so we focus on $(\omega,2\omega)$ scheme with the trapezoidal pulse shape in this paper.

\nopagebreak
\section{Theory}
\label{sec:Model}

The rotational dynamics of the linear polar molecule is studied in the presence of the nonresonant linearly polarized laser pulses. The molecule is assumed to be in its ground vibronic state. The molecular dynamics is thus treated within the rigid-rotor approximation (frozen internal vibrational motion), interacting with the laser fields. The dynamics of the system is described by the time-dependent Shr\"{o}dinger equation:
\begin{equation}\label{eqSE}
i\hbar \frac{d}{dt}\psi(\theta,\phi,t) = H(\theta,t)\psi(\theta,\phi,t)\,.
\end{equation}
where the Hamiltonian is written as
\begin{equation}\label{eqHamiltonian}
H(\theta,t)=BJ^2+V(\theta,t).
\end{equation}
with $\psi(\theta,\phi,t)$ being the time-dependent wave function, $\theta,\phi$~--- the Euler angles, $B=\hbar/(4\pi I c)$~ --- the rotational constant ($c$ is the speed of light and $I$ is the molecular moment of inertia) and $J^2$~--- the squared angular momentum operator. The first term in equation~(\ref{eqHamiltonian}) is the rotational energy of the molecule. The interaction potential $V(\theta,t)$ of the molecule with the laser field $E(t)$ is written as~\cite{kanai2001numerical}
\begin{align}\label{eqPotential}
    V(\theta,t)&= -\mu_{0}E(t)\cos(\theta) + W_{pol}(\theta,t) + W_{hyp}(\theta,t),
\end{align}
where $\mu_{0}$ denotes the molecule's permanent dipole moment. The interaction potential with polarizability $W_{pol}(\theta,t)$ is defined as:
\begin{align}\label{eqPotentialPolarizabilityPart}
    W_{pol}(\theta,t) = &-\tfrac{1}{2} E^2(t) \left[ (\alpha_\parallel-\alpha_\perp)\cos^2(\theta) + \alpha_\perp\right],
\end{align}    
and the interaction potential with the hyperpolarizability $W_{hyp}(\theta,t)$ has the form:
\begin{align}\label{eqPotentialHyperpolarizabilityPart}
    W_{hyp}(\theta,t) = &-\tfrac{1}{6} E^3(t) \left[ (\beta_\parallel-3\beta_\perp)\cos^3(\theta) + 3\beta_\perp\cos(\theta)\right].
\end{align}
Here $\theta$ is the polar angle between the internuclear axis and the direction of the field polarization. It is precisely this angle which defines the alignment and the orientation of the molecule with respect to the laser field.  The terms $\alpha_\parallel$ and $\alpha_\perp$ ($\beta_\parallel$ and $\beta_\perp$) correspond to the polarizability (hyperpolarizability) components, respectively, parallel and perpendicular to the molecular axis. 

The linearly polarized laser field $E(t)$ is defined as
\begin{align}\label{eqElectricFieldDescription}
    E(t)=&\left[E_{0}f_1(t)\gamma_1 \cos(\omega t)+ \right.\notag\\
    &\left.E_{0}f_1(t)\sqrt{(1-\gamma_1^2)}\cos(2\omega t+\delta^{CEP}_1) \right] +\notag\\
    &\left[E_{0}f_2(t-t_{d})\gamma_2 \cos(\omega (t-t_{d}))+\right. \notag\\
    &\left.E_{0}f_2(t-t_{d})\sqrt{(1-\gamma_2^2)}\cos(2\omega (t-t_{d})+\delta^{CEP}_2) \right]
 \end{align}
Here $\omega$ is the pulse frequency; $\delta^{CEP}_j$ is the relative carrier-envelope phase (CEP) of the second harmonic with respect to the fundamental of the $j$-th pulse ($j=1,2$); $t_d$ is the time delay between the first and second pulses; $E_{0}$ is the electric-field peak amplitude. 
The parameter $\gamma_j$ (${0 \leq \gamma_j \leq 1}$) defines the relative strength of the $\omega$ and the $2\omega$ components within the $j$-th pulse ($j=1,2$): the pulse is monochromatic at ${\gamma_j = 0, 1}$ and the peak amplitudes of the fundamental and second harmonic are equal each other at ${\gamma_j^2 = 1/2}$. This definition of the parameter $\gamma_j$ allows keep the total average intensity constant and proportional to $E_0^2$ when $\gamma_j$ varies~\cite{tehini2008field}. Thus, the peak intensities of the 
fundamental and second harmonics are  
${I^{(\omega)}_{0}=E_{0}^2\gamma_j^2}$ and ${I^{(2\omega)}_{0}=E_{0}^2(1-\gamma_j^2)}$, 
respectively. 
The total peak intensity of each pulse ${I_{tot}=I_{0}^{(\omega)} + I_{0}^{(2\omega)}=E_{0}^2}$.

The laser pulse envelopes ($|f_j(t)|\leq 1$) of the fundamental and second harmonic of the $j$th pulse are written as:
\begin{equation}
f_j(t) = 
\left\{
    \begin{array}{lll}
        0,              & \text{if } &t \leq t_0 \\ 
        2.5 + 4t/\tau_{j},  & \text{if } &t_0 < t < t_1\\
        1,              & \text{if } &t_1 < t < t_2\\ 
        2.5 - 4t/\tau_{j},  & \text{if } &t_2 < t < t_3\\
        0,              & \text{if } &t \geq t_3
    \end{array}
\right.
\end{equation}
where $\tau_{j}$ is the  full width at half maximum (FWHM), ${t_2-t_1}$ is the trapezoidal plateau time, ${t_1-t_0}$ is the risetime (${t_3-t_2}$ --- the falltime) of $j$th pulse. 
We use the trapezoidal shape with the plateau time doubled with respect to the risetime and falltime ${{t_2-t_1} = 2({t_1-t_0}) = 2({t_3-t_2})}$.

The field-dressed rotational dynamics is analyzed in terms of the expectation values:
\begin{equation}
    \langle \cos^k(\theta) \rangle_{T=0}=\iint
    \cos^k(\theta)|\psi(\theta,t)|^2  \sin(\theta)d\theta d\phi, k=1,2.
\end{equation}
For the orientation and the alignment $k=1$~and~$2$, respectively.
For the of the molecules in the absence of external fields the degree of alignment $\langle \cos^2(\theta) \rangle \in [0,1]$ is equal $1/3$ for the isotropic distribution evenly distributed across all $\theta$. For perfect alignment $\langle \cos^2(\theta) \rangle \simeq 1$, whereas ${\langle \cos^2(\theta) \rangle = 0}$ for the ``anti-alignment'' (the molecular axes are perpendicular to the electric field). By definition of the degree of orientation $\langle \cos(\theta) \rangle$ its different expectation values belong to the interval $[-1,1]$. A molecule is said to be oriented along a given axis of the space-fixed frame if one of the expectation values satisfies $|\langle \cos(\theta) \rangle| \simeq 1$. 

A system of molecular gas interacting with the laser pulse can be assumed to be a thermal ensemble characterized by the temperature $T$. Hence, the non-zero temperature is accounted by averaging the alignment and the orientation parameters over the ensemble with the Boltzmann distribution of angular momentum among various molecules in the gas~\cite{machholm2001postpulse,tehini2012field}: 
\begin{align}\label{eqAlignmentOrientationDefinitionTemperatureAveraged}
    &\langle \cos^k(\theta) \rangle_{T \neq 0}=\sum_{J_i}^{\infty} \tfrac{1}{\mathcal{Z}}g_{J_i}\exp\left[-B J_i(J_i+1)/k_{B}T\right] \times \notag\\ 
    &\sum_{M_i=-{J_i}}^{J_i} \langle \psi_{J_i,M_i}(t)| \cos^k(\theta)  |\psi_{J_i,M_i}(t)\rangle, k=1,2
\end{align}
where $k_B$ denotes the Boltzmann constant, the term $g_{J_i}$ is the nuclear spin degeneracy (is equal to 1 for heteronuclear molecules), and ${\mathcal{Z} = \sum_{J_i=0}^{\infty} (2J_i+1)g_{J_i}\exp\left[-B J_i(J_i+1)/k_{B}T\right]}$ --- the rotational partition function.
The initial distribution of rotational levels is given by the Boltzmann distribution.
The molecular alignment induced by the laser pulse is the averaged effect of the different rotational wave packets formed from each
Boltzmann-weighted initial rotational state.

The rotational wave function is expanded using the finite basis set of $|J,M\rangle$, which is a set of the eigenfunctions of a rigid rotor in a field-free space. Because we defined the polarization axes of both the pulses as the laboratory-fixed $Z$ axis, the Hamiltonian is cylindrically symmetric about the $Z$ axis. Therefore, $M$ becomes a good quantum number, that is, $\Delta M = 0$. 

The time-dependent Shr\"{o}dinger equation~(\ref{eqSE}) is solved by the a split operator method~\cite{marchuk1990splitting}. Hereafter, we use HBr molecule as an illustrative example. Numerical values of
the molecular parameters are taken as $B~=~8.3482$~cm$^{-1}$, $\mu_0~=~0.828~$D, $\alpha_\parallel=3.64~$\AA$^3$,  $\alpha_\perp=3.315~$\AA$^3$, $\beta_\parallel=-1.07\cdot10^9~$\AA$^5$, $\beta_\perp=4.3\cdot10^8~$\AA$^5$~\cite{bishop1999effects}. 
The rotational period of the HBr molecule can be calculated $T_{rot}=1/(2Bc) \approx 1.998$~ps, where $c$ is the speed of light.
The laser pulse frequency is taken to be~$12500~$cm$^{-1}$ (corresponding to the laser wavelength~$800~$nm). 
The total intensity $I_{tot}$ of each pulse was set to ${7\cdot10^{13}}$~W$/$cm$^{2}$ if not otherwise indicated.

\section{Results}
\label{sec:Results}

\subsection{One two-color trapezoidal pulse}
\begin{figure}[tbp]
\begin{overpic}
{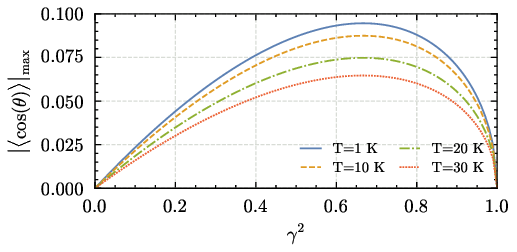}
\end{overpic}
\caption{\label{figOnePulseTwoColorOrientationRelativeIntensity}The dependence of the maximum degree of orientation ($|\langle\cos(\theta)\rangle|_{\max}$) on the squared relative strength $\gamma^2$ of the fundamental and second harmonic intensities for the single two-color ${0.12}$~ps pulse. The different temperatures $T = 1, 10, 20, 30$~K are indicated by a solid, dashed, dash-dotted and dotted lines respectively.} 
\end{figure}

\begin{figure*}[!htbp]\centering
\begin{overpic}[width=\textwidth]{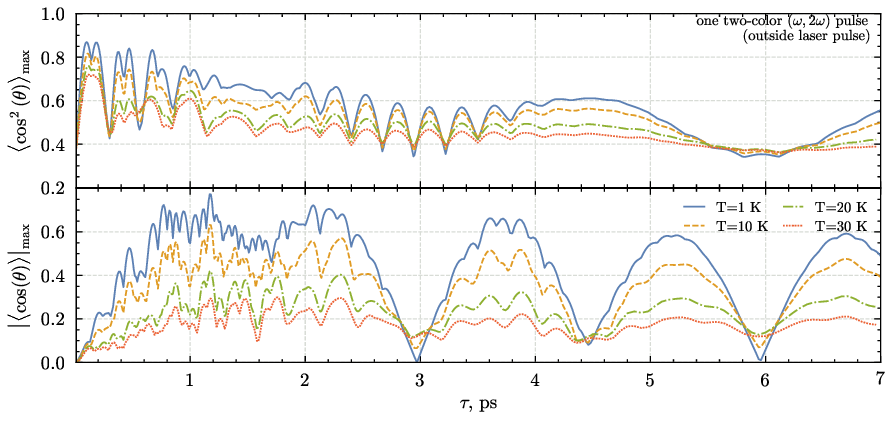}
\put(9.5,44){(a)}
\put(9.5,24){(b)}
\end{overpic}
\caption{The dependence of the maximum the degrees of alignment ($\langle\cos^2(\theta)\rangle_{\max}$) (a) and orientation ($|\langle\cos(\theta)\rangle|_{\max}$) (b)  (outside the laser pulse) on the pulse duration $\tau$ at the total intensity ${7\cdot10^{13}}$~W$/$cm$^{2}$. The different temperatures $T = 1, 10, 20, 30$~K are indicated by a solid, dashed, dash-dotted and dotted lines respectively.} 
\label{figOnePulseTwoColorAlignemntAndOrientationOutsidePulse}
\end{figure*}
\begin{figure}[!htbp]\centering
\begin{overpic}
{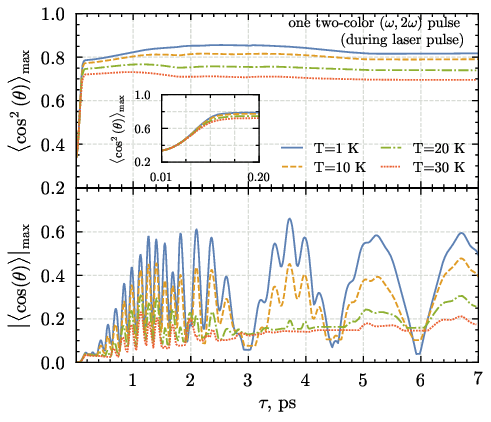}
\put(16.5,79){(a)}
\put(16.5,44){(b)}
\end{overpic}
\caption{The dependence of the maximum alignment ($\langle\cos^2(\theta)\rangle_{\max}$) (a) and orientation ($|\langle\cos(\theta)\rangle|_{\max}$) (b) degrees (during the pulse time) on the pulse duration $\tau$ at the total intensity $7\cdot10^{13}$~W$/$cm$^{2}$. The inset shows a magnified region near the origin. The different temperatures $T = 1, 10, 20, 30$~K are indicated by a solid, dashed, dash-dotted and dotted lines respectively.} 
\label{figOnePulseTwoColorAlignemntAndOrientationDuringPulse}
\end{figure}
\begin{figure}[htb]\centering
\begin{overpic}{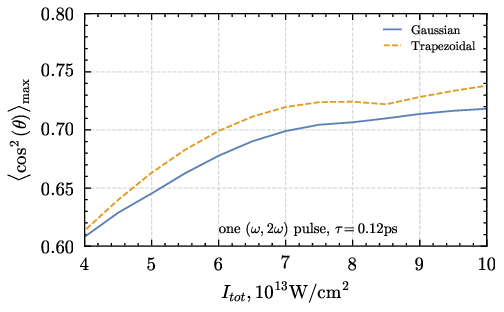}
\end{overpic}
\caption{The maximum degree of the molecular alignment $\langle\cos^2(\theta)\rangle_{\max}$ dependence on the intensity of the two-color $0.12$~ps laser pulse with the trapezoidal shape of pulse (solid line) and the conventional Gaussian pulse shape (dashed line) at the temperature $30$~K.} 
\label{figOnePulseTwoColorIntensityCompareWithGauss}
\end{figure} 

The molecular alignment and orientation induced by the one two-color trapezoidal laser pulse were investigated. For simplicity, we have omitted the index with the pulse no. and all quantities of Eq.~(\ref{eqPotentialHyperpolarizabilityPart}) in this subsection are assumed to relate to the first pulse. 

The \textit{optimal} value of the varying parameter is defined as the value at which the maximum alignment degree or the orientation degree is reached, while the other parameters are kept fixed in simulations.

First, we studied how the relative strength $\gamma$ of the fundamental and second harmonic intensities affects the degree of the orientation. The calculated dependence of the maximum degree of orientation on the relative strength $\gamma$ of the fundamental and second harmonics is illustrated in Fig.~\ref{figOnePulseTwoColorOrientationRelativeIntensity}. The analysis of the Fig.~\ref{figOnePulseTwoColorOrientationRelativeIntensity} shows that the maximum degree of orientation is reached at ${\gamma^2=2/3 \approx 0.67}$, that coincides with the known results for the Gaussian pulses~\cite{tehini2008field}. Thus, the optimal ratio of the fundamental and second harmonic intensities  $I_{0}^{(\omega)}/I_{0}^{(2\omega)}=2$ is used hereafter for two-color pulse unless otherwise stated.

In order to find the optimal two-color pulse duration for the higher degree of alignment (orientation), we have calculated the maximum values of the degree of alignment, obtained outside and during the pulse, and illustrated the obtained dependencies in Fig.~\ref{figOnePulseTwoColorAlignemntAndOrientationOutsidePulse}(a) (Fig.~\ref{figOnePulseTwoColorAlignemntAndOrientationOutsidePulse}(b)) and in Fig.~\ref{figOnePulseTwoColorAlignemntAndOrientationDuringPulse}(a) (Fig.~\ref{figOnePulseTwoColorAlignemntAndOrientationDuringPulse}(b)) respectively. 
The inset in Fig.~\ref{figOnePulseTwoColorAlignemntAndOrientationDuringPulse}(a) shows a magnified region near the origin.

The alignment degree maximum \textit{(during the pulse)}, depicted in Fig.~\ref{figOnePulseTwoColorAlignemntAndOrientationDuringPulse}(a), reaches its saturation at $0.2$~ps, and then it remains almost constant regardless of the pulse duration, that is shown in the inset in Fig.~\ref{figOnePulseTwoColorAlignemntAndOrientationDuringPulse}(a). Whereas the dependency of the maximum  \textit{(outside the pulse)} degree of alignment on the duration shows the clear periodic structures with different periods of oscillations. For the $\tau<4$~ps~$ = 2T_{rot}$ the character period of such oscillations is $\approx0.275$~ps~$ =0.1375 T_{rot}$.
Whereas for the adiabatic regime of the long pulses $\tau>4$~ps~$ =2T_{rot}$ the maximum alignment degree's dependency on $\tau$ has a pronounced oscillations with period $\approx3$ps~$ =1.5T_{rot}$.

For the temperature $30$~K the maximum alignment degree  $\approx0.72$ (outside the pulse) is reached at the $\tau\approx 0.12$~ps. Whereas for the temperature $1$~K the maximum alignment degree $\approx0.867$ (outside the pulse) is reached at the $\tau\approx 0.18$~ps. This value of the maximum alignment degree after \textit{one two-color trapezoidal pulse} is as high as the alignment degree maximum, attained by the train of \textit{six one-color pulses} (with square or $sin^2$ shapes) with the comparable intensity and duration for the same HBr molecule at zero temperature in Ref.\cite{tyagi2017effect}. 
Thus, the trapezoidal shape of pulses is shown to be more efficient than $sin^2$ and square pulses.

The analysis of Figs.~\ref{figOnePulseTwoColorAlignemntAndOrientationOutsidePulse}(b) and~\ref{figOnePulseTwoColorAlignemntAndOrientationDuringPulse}(b) shows, that the longer pulse duration leads to a significant increase in the orientation degree. 
For the short two-color pulses $\tau<0.2$~ps the orientation degree for non-zero temperatures $T>20$~K is by an order of magnitude smaller than for the cold temperature $T=1$~K.

\begin{figure}[!htbp]
        \begin{overpic}{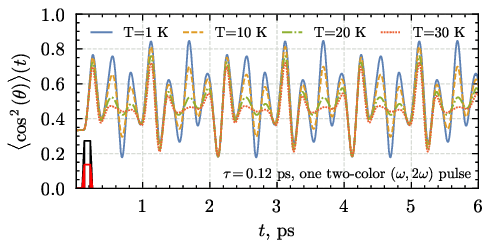}
        \put(39,14.5){(a)}
        \end{overpic}
        \vspace{-.5cm}
        \begin{overpic}{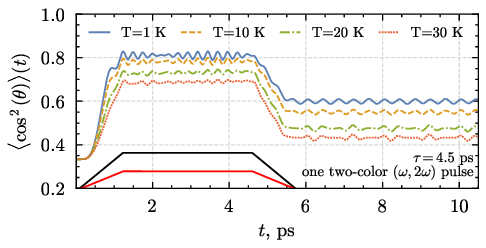}
        \put(78,18){(b)}
        \end{overpic}
    \caption{\label{figOnePulseTwoColorAlignmentTimeDependence}The time evolution of the alignment degree $\langle\cos^2(\theta)\rangle(t)$ after the one two-color laser pulse with the $0.12$~ps~(a) and $4.5$~ps~(b) duration. 
        The different temperatures $T = 1, 10, 20, 30$~K are indicated by a solid, dashed, dash-dotted and dotted lines respectively. The temporal shape (arbitrary unit for height) of the two-color pulse is plotted at the bottom of each panel (in black for the fundamental and in red or grey for the second harmonic).} 
        \begin{overpic}{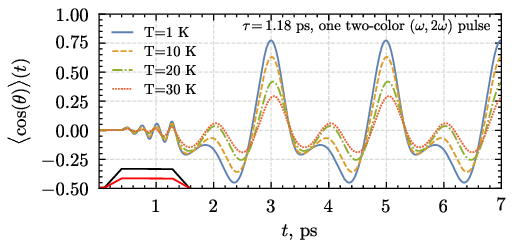}
        \put(92,13){(a)}
        \end{overpic}
        \begin{overpic}{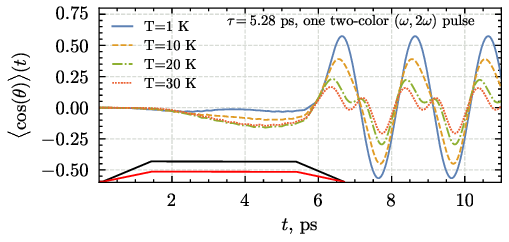}
        \put(92,14){(b)}
        \end{overpic}
        \vspace{-.5cm}
    \caption{\label{figOnePulseTwoColorOrientationTimeDependence}The time evolution of the degree of orientation $\langle\cos(\theta)\rangle(t)$ after the one two-color laser pulse with the $1.18$~ps~(a) and $5.28$~ps~(b) duration.
        The different temperatures $T = 1, 10, 20, 30$~K are indicated by a solid, dashed, dash-dotted and dotted lines respectively. The temporal shape (arbitrary unit for height) of the two-color pulse is plotted at the bottom of each panel (in black for the fundamental and in red or grey for the second harmonic).} 
\end{figure} 

For cold temperatures T~$=1$~K, the maximum degree of orientation is greatly enhanced in the pulse duration interval $0.4$~ps~${=0.2T_{rot}\leq\tau\leq2.6}$~ps~$=1.3T_{rot}$ reaching its peak value $\approx0.774$ at $\tau=1.17$~ps. 
For the adiabatic regime of the long pulses ${\tau>2.5}$~ps~${=1.25T_{rot}}$ the maximum orientation degree dependence on  $\tau$ has a pronounced oscillations with period $\approx3T_{rot}/4$.
The above-mentioned oscillations of dependencies of the maximum alignment and orientation degrees, that were not reported before by other authors (for other shapes of laser pulses) to the best of our knowledge, do not vanish for non-zero temperatures, but their amplitudes evidently decrease with the temperature increase.

We also carried out the comparison of the efficiency of the trapezoidal pulses with the conventional Gaussian pulse on the achieved maximum degree of the alignment. The corresponding maximum alignment degrees as functions of the pulse intensity for the Gaussian (solid line) and the trapezoidal (dashed line) pulse shapes are shown in Fig.~\ref{figOnePulseTwoColorIntensityCompareWithGauss}. 
The trapezoidal pulse generates the higher alignment than the Gaussian pulse of the same intensity and the duration, which is defined as the full width at half maximum for both shapes of pulse.

Figure~\ref{figOnePulseTwoColorAlignmentTimeDependence} depicts the time evolution of the degree of molecular alignment created by the one trapezoidal laser pulse for the different rotational temperatures $T = 1,10,20,30$~K, marked by the solid, dashed, dash-dotted and dotted lines respectively, at the laser total peak intensity $7\cdot10^{13}$~W/cm$^{2}$. The pulse duration is $\tau = 0.12~$ps corresponding to the non-adiabatic regime of the molecular alignment in Fig.~\ref{figOnePulseTwoColorAlignmentTimeDependence}(a), whereas the time-dependence of the alignment degree at $\tau = 4.5~$ps is shown in Fig.~\ref{figOnePulseTwoColorAlignmentTimeDependence}(b). We note that the trapezoidal shape of the pulse causes after-pulse alignment even in adiabatic regime $\tau>T_{rot}$, in contrast with the commonly used Gaussian shape where the molecular alignment after pulse vanishes in adiabatic regime~\cite{torres2005dynamics}.

The time evolution of the orientation degree are illustrated in Figs.~\ref{figOnePulseTwoColorOrientationTimeDependence} for the laser pulses with the optimal durations: $\tau = 1.18~$ps (a) for the non-adiabatic regime and $\tau = 5.28~$ps (b) for the adiabatic regime. The different temperatures $T = 1, 10, 20, 30$~K are indicated by a solid, dashed, dash-dotted and dotted lines respectively. 

\begin{figure}[H]
\begin{overpic}
{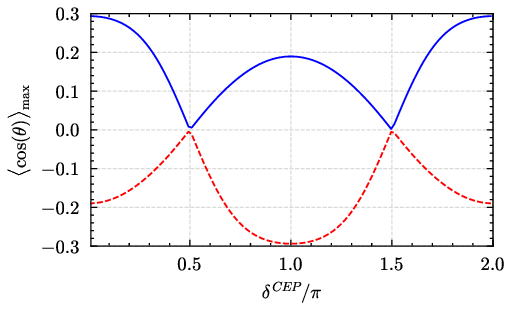}
\end{overpic}
\caption{\label{figOnePulseTwoColorOrientationRelativePhase}The maximum degrees of the positive (solid line) and negative orientations (dashed line) ($\langle\cos(\theta)\rangle_{\max}$) as a function of the relative phase $\delta^{CEP}\equiv\delta^{CEP}_1$ between the fundamental field and its second harmonic field of the pulse at T$=30$K. The laser parameters are $\tau=1.18$~ps, $I_{tot}=7\cdot10^{13}$~W/cm$^2$. 
} 
\vspace{-0.5cm}
\end{figure}

The effect of varying relative phase on the alignment and orientation degrees was further considered.
The averaged over the fast oscillations of the laser field~\cite{tehini2012field,koch2019quantum} polarizability part of interaction $W_{pol}(\theta,t)$ does not contain any dependence on the relative phase $\delta^{CEP}$. Thus the degree of alignment, that is mainly sensitive to $W_{pol}(\theta,t)$, is not almost affected by the relative phase.
Whereas the preferred orientation direction is changed from positive to negative by the varying {$\delta^{CEP}$ from $0$ to $\pi$} for the one two-color trapezoidal pulse, considered in this subsection.
The influence of the relative phase $\delta^{CEP}$ on the orientation degree is illustrated in Fig.~\ref{figOnePulseTwoColorOrientationRelativePhase}.
Averaging over the fast oscillations of the laser field~\cite{tehini2012field,koch2019quantum} of the hyperpolarizability part of interaction leads to the factor $\cos(\delta^{CEP})$ in the ${W_{hyp}(\theta,t)}$, that explains the orientation degree vanishing at the points $\delta^{CEP}=\pi/2,3\pi/2$  and the presence of the symmetry of the obtained curve about the line $\delta^{CEP}=\pi$ in Fig.~\ref{figOnePulseTwoColorOrientationRelativePhase}. 
Note that the maximum of the positive orientation degree at point $\delta^{CEP}=0$ have the same absolute value as the negative orientation degree at point $\delta^{CEP}=\pi$.
Thus the relative phase allows to easily control the desired preferred direction of the molecules' axes orientation during after pulse dynamics.

\subsection{Hybrid strategy}

We also tested the hybrid strategy, proposed by Zhang~\cite{zhang2011controlling} for the Gaussian profile of the laser pulse and experimentally proven in Ref.~\cite{ren2014alignment}, where the two-color pulse is preceeded by the one-color pulse in order to enhance the orientation by the adjusting the time delay between the pulses. The optimal delay was found close to $T_{rot}/4$ or $3T_{rot}/4$ for molecules with a small hyperpolarizability~\cite{zhang2011controlling,tehini2012field}. 
Tehini~\cite{tehini2012field} showed that for molecules with a large hyperpolarizability, a~monochromatic laser pre-pulse should be applied before the two-color laser pulse at a time close to the rotational period $T_{rot}$.

\begin{figure}[t]
\begin{overpic}
{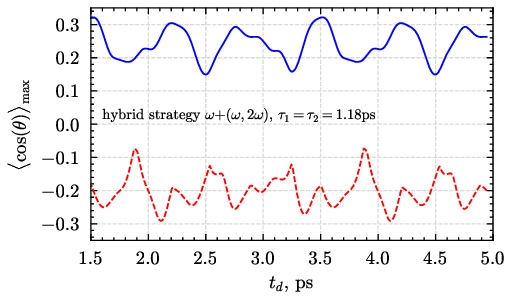}
\put(89,14){(a)}
\end{overpic}
\begin{overpic}
{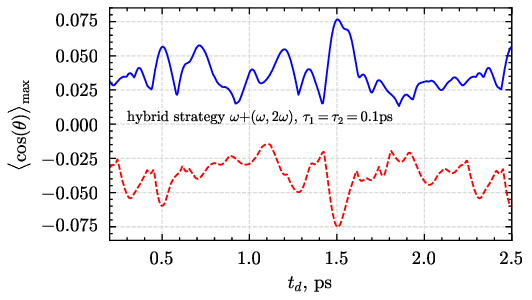}
\put(89,14){(b)}
\end{overpic}
\caption{\label{figTwoPulsesBipulseStrategy} The dependence of the maximum degrees of the positive (solid line) and negative orientations (dashed line) ($\langle\cos(\theta)\rangle_{\max}$) on the time delay $t_d$ between the monochromatic and two-color pulses for the optimal $\tau_1=\tau_2=1.18$~ps (a) and not optimal $\tau_1=\tau_2=0.1$~ps (b) laser pulse durations. The laser pulse parameters are
$I_{tot}=7\cdot10^{13}$~W/cm$^2$, $T=30$~K. 
} 
\end{figure}

First, we proved whether this approach would give any advantages for the two-color pulse with $1.18$~ps duration, that is optimal to obtain higher degree of orientation. The dependency of the orientation degree on the pulse delay between monochromatic and two-color pulses is depicted in Fig.\ref{figTwoPulsesBipulseStrategy}(a). Analysis of the Fig.\ref{figTwoPulsesBipulseStrategy}(a) shows that there is no additional increase in the maximum orientation degree in comparison with the single two-color pulse with the optimal $1.18$~ps duration. 

Second, we applied this approach to the system by the  monochromatic and two-color pulses when both pulses were short with $0.1$~ps duration. That two-color pulse's duration is not optimal for reaching the maximum degree of orientation by a single pulse as clearly seen on Fig.~\ref{figOnePulseTwoColorAlignemntAndOrientationOutsidePulse}(b). The maximum degree of orientation is equal $0.057$ for the single two-color $0.1$~ps pulse at $T=30$~K and $I_{tot}=7\cdot10^{13}$~W/cm$^2$. We were interested in how much the use of the hybrid strategy would enhance the degree orientation by adjusting the time delay between monochromatic and two-color trapezoidal pulses with not optimal duration. The obtained results are illustrated in Fig.\ref{figTwoPulsesBipulseStrategy}(b), where maximum degree of orientation dependency on the time delay $t_d$ between the monochromatic prepulse and two-color pulse is shown. Applying monochromatic prepulse with $3T_{rot}/4=1.5$~ps allows to increase the maximum orientation degree $1.3$ times with comparison of the single two-color $0.1$~ps trapezoidal pulse.

Thus, the hybrid strategy with the combination of monochromatic prepulse and two-color pulse has clear advantages for short pulses with not-optimal durations. Instead, for pulses with the optimal duration such a monochromatic prepulse is likely to reduce the maximum orientation degree.

\begin{figure}[tb]
\begin{overpic}
{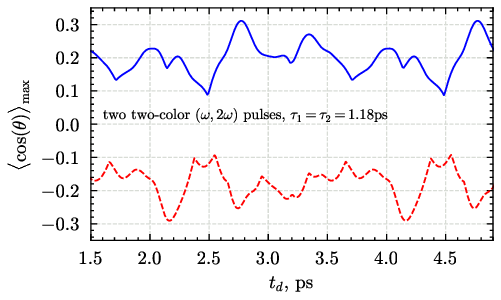}
\put(92,14){(a)}
\end{overpic}
\begin{overpic}
{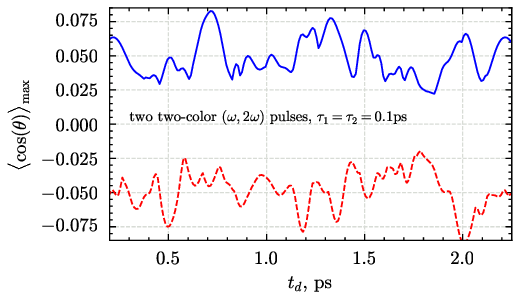}
\put(92,13){(b)}
\end{overpic}
\caption{\label{figTwoPulsesTwoColorTimeDelay}The dependence of the maximum degrees of the positive (solid line) and negative orientations (dashed line) ($\langle\cos(\theta)\rangle_{\max}$) on the time delay $t_d$ between first and second two-color pulses for the optimal $\tau_1=\tau_2=1.18$~ps~(a) and not optimal $\tau_1=\tau_2=0.1$~ps~(b) laser pulse durations. The laser parameters are $I_{tot}=7\cdot10^{13}$~W/cm$^2$, $T=30$~K.
}
\end{figure}

\subsection{
Two two-color trapezoidal pulses
}

We also wondered whether the application of the scheme with a two-color prepulse~\cite{wu2010field} is preferrable to increase the maximum orientation degree for trapezoidal shape of pulse considered in the present paper.
In order to compare the results of the hybrid strategy~\cite{zhang2011controlling} and of the use of the two-color prepulse~\cite{wu2010field}, the calculations with same field parameters as in previous subsection were carried out. 

\begin{figure}[!htbp]\centering
\begin{overpic}
{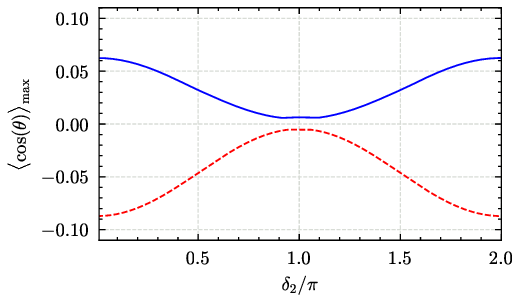}
\end{overpic}
\caption{\label{figTwoPulsesTwoColor100fs_Phase}The maximum degrees of the positive (solid line) and negative orientations (dashed line) $\langle\cos(\theta)\rangle_{\max}$ as a function of the relative phase $\delta^{CEP}_2$ between the fundamental field and its second harmonic field of the second pulse, that is applied at $t_d = 2.0$~ps~$=T_{rot}$. The laser parameters are $\tau_1=\tau_2=0.1$~ps, $\delta^{CEP}_1=0$, 
$I_{tot}=7\cdot10^{13}$~W/cm$^2$.
} 
\begin{overpic}
{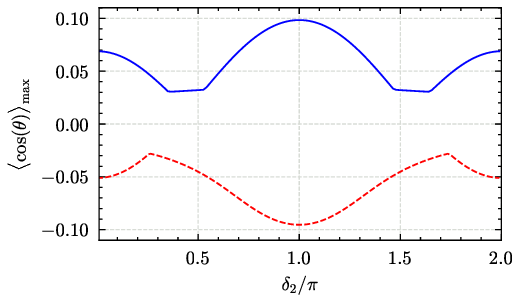}
\end{overpic}
\caption{\label{figTwoPulsesTwoColor100fs_PhaseTimeDelay_1500fs}The dependence of the maximum degrees of the positive (solid line) and negative orientations (dashed line) $\langle\cos(\theta)\rangle_{\max}$ on the relative phase $\delta^{CEP}_2$ ($\delta^{CEP}_1=0$) between the fundamental field and its second harmonic field of the second pulse, that is applied at $t_d = 1.5$~ps~$\approx 3 T_{rot}/4$. The laser parameters are $\tau_1=\tau_2=0.1$~ps, $\delta^{CEP}_1=0$, 
$I_{tot}=7\cdot10^{13}$~W/cm$^2$.
} 

\end{figure}

The dependency of the orientation degree on the pulse delay between two $1.18$~ps two-color pulses is depicted in Fig.\ref{figTwoPulsesTwoColorTimeDelay}(a). Thus, just as for the hybrid strategy, 
the two-color prepulse does not give any additional gain in the maximum orientation degree in the case of the pulses with the optimal duration $1.18$~ps. On the contrary, the two-color prepulse is likely to destroy maximum orientation degree obtained by single two-color pulse with the optimal duration. 

The obtained results for two short $0.1$~ps two-color pulses are shown in Fig.~\ref{figTwoPulsesTwoColorTimeDelay}(b), revealing that two-color prepulse with $t_d=T_{rot}$ produces a slightly higher value of the maximum orientation degree than the adding of the monochromatic prepulse with $t_d=3T_{rot}/4$ for the hybrid strategy. There are also additional points of time delay between two pulses, where the maximum orientation degree is increased in comparison with the effect of the single pulse. 

We also investigated how varying of the relative phase $\delta^{CEP}_2$ of the second pulse ($\delta^{CEP}_1=0$) affect the orientation degree for two short $0.1$~ps two-color pulses with the optimal time delay $t_d=T_{rot}=2$~ps. 
It is shown in Fig.~\ref{figTwoPulsesTwoColor100fs_Phase}, that although the further increase of the maximum orientation degree is not possible by the relative phase in this case, there is an unique possibility to ``turn off'' (substantially suppress) the orientation created by the first pulse with the help of the second pulse, applied at $nT_{rot} (n\in Z^+)$ with the same parameters, but $\delta^{CEP}_2=\pi$. This effect coincides with the known results of the orientation degree control by the relative phase for the two two-color Gaussian pulses~\cite{wu2010field}.

While the relative phase varying is not profitable for the pulses with the optimal time-delay, it could be promising in the case of two pulses with the not fine-tuned delay. The calculated maximum orientation degree as a function of the relative phase $\delta^{CEP}_2$ of second pulse is illustrated in Fig.~\ref{figTwoPulsesTwoColor100fs_PhaseTimeDelay_1500fs}. The strong enhancement of the orientation degree $1.4$~times is seen at $\delta^{CEP}_2=\pi$. 

\section{Conclusion}
\label{sec:Conclusions}

We have theoretically investigated the field-free alignment and orientation of the linear molecules by two-color trapezoidal laser pulses. Our results showed that the trapezoidal laser pulse allows to enhance the maximum alignment degree for the same intensity and duration comparing to the conventional Gaussian laser pulse. The alignment and orientation persist after the pulse for both non-adiabatic and adiabatic regimes. While the maximum (during the pulse) alignment degree quickly saturates and remains almost constant with the pulse duration increase, the dependencies of the maximum (outside the laser pulse) alignment and orientation degrees on the pulse duration show the clear periodic structures in the adiabatic regime. The effect of the non-zero temperature is also shown. The variation of the relative phase between the fundamental and second harmonic of the single pulse allows to control the desired preferred direction of the molecules' axes orientation during after pulse dynamics. We also compared the use of two two-color pulses with the hybrid strategy (where a monochromatic prepulse is applied) for maximizing the degree of orientation at non-zero temperature. Neither the two two-color pulses, nor the hybrid strategy are useful for the pulses with the optimal duration. Whereas for the short pulses the application of the two-color prepulse at the time delay $T_{rot}$ increases the maximum orientation degree $1.54$ times, whereas the hybrid strategy's gain (with the monochromatic prepulse at the time delay $3T_{rot}/4$) is slightly smaller with the increase factor $1.34$. For the two two-color pulses technique the relative phase variation of the second two-color pulse (applied with the time delay $T_{rot}$) allows to substantially suppress the orientation degree, similar to the known~\cite{wu2010field} orientation ``turn-off'' effect for Gaussian pulses, whereas for the two two-color pulses with not fine-tuned delay it could increase the maximum orientation degree.
 

%

\end{document}